# Revealing mode formation in quasi-bound states in the continuum metasurfaces via near-field optical microscopy

T. Gölz[1,†], E. Baù[1,†], A. Aigner[1], A. Mancini[1,2], M. Barkey[1], F. Keilmann[1], S. A. Maier[3,4], A. Tittl[1*]

1. Chair in Hybrid Nanosystems, Nanoinstitute Munich and Center for Nanoscience (CeNS), Faculty of Physics, Ludwig-Maximilians-Universität München, 80539 Munich, Germany.

2. Centre for Nano Science and Technology, Fondazione Istituto Italiano di Tecnologia, Via Rubattino 81, Milan, 20134 Italy

3. School of Physics and Astronomy, Monash University, Clayton, Victoria 3800, AUS

4. Department of Physics, Imperial College London, London SW7 2AZ, U.K.

†These authors contributed equally

Email: andreas.tittl@physik.uni-muenchen.de

**Abstract**

Photonic metasurfaces offer exceptional control over light at the nanoscale, facilitating applications spanning from biosensing, and nonlinear optics to photocatalysis. Many metasurfaces, especially resonant ones, rely on periodicity for the collective mode to form, which makes them subject to the influences of finite size effects, defects, and edge effects, all of which have considerable negative impact at the application level These aspects are especially important for quasi-bound state in the continuum (BIC) metasurfaces, for which the collective mode is highly sensitive to perturbations due to high-quality factors and strong near-field enhancement. Here, we quantitatively investigate the mode formation in quasi-BIC metasurfaces on the individual resonator level using scattering scanning near-field optical microscopy (s-SNOM) in combination with a new image processing technique. We find that the quasi-BIC mode is formed at a minimum size of 10 x 10-unit cells much smaller than expected from far-field measurements. Furthermore, we show that the coupling direction of the resonators, defects and edge states have pronounced influence on the quasi-BIC mode. This study serves as a link between the far-field and near-field responses of metasurfaces, offering crucial insights for optimizing spatial footprint and active area, holding promise for augmenting applications such as catalysis and biospectroscopy.

**Introduction**

A fundamental goal in photonics is the confinement of light within subwavelength volumes to drastically enhance light-matter interactions. Such increased interactions unlock unique applications in diverse fields such as lasing,[1,2] nonlinear optics,[3–5] structural color[6,7], quantum photonics,[8,9] ultrasensitive biosensing[10,11], enhanced chiral sensing[12–14] and photocatalysis.[15,16] A pivotal platform technology for nanophotonics are optical metasurfaces comprising two-dimensional arrays of nanostructured elements with dimensions smaller than the wavelength of light, which excel in finely tuning the interaction of light with matter.[17–19]

In particular, all-dielectric metasurfaces supporting photonic bound states in the continuum (BICs) are a promising new platform for the precise engineering of strong light-matter interactions.[20–23] Due to the geometry of the subwavelength resonators, BICs are "dark states" that cannot couple to the far-field and thus have an infinite quality factor (Q factor, defined as the resonance wavelength divided by the line width). Small perturbations can transform these "dark" BIC modes into quasi-BIC states, resulting in very high but finite Q factors that are accessible from the far field.[21] For nanophotonic resonators in metasurfaces, the class of symmetry protected BICs is particularly advantageous as the breaking of the symmetry can be introduced in the manufacturing process through geometrical asymmetries in the resonator structures.[20] This concept gives rise to all-dielectric metasurfaces with high and easily controllable quality factors leading to high near-field enhancements[20]. Moreover, the spectral position of the metasurface resonances exhibits flexible tunability across a broad frequency spectrum, facilitated by the scaling of the resonator dimensions, while control over the Q factor adjustability is achieved through the precise tuning of resonator asymmetry parameters.[24]

However, to date, almost all studies on quasi-BIC metasurfaces rely on investigating quasi-BICs via far-field spectroscopy techniques.[4] While providing valuable insights, such approaches offer only limited information on the mode properties of metasurfaces. In comparison, investigations into the near-field characteristics of the collective quasi-BIC mode, important for a fundamental understanding of photonic modes and pivotal for numerous applications such as biosensing or catalysis[16], remain notably sparse. Crucial properties concerning the minimum array size for complete mode formation, the directionality of the coupling between unit cells, and the effect of the edge of a metasurface and defects on the mode formation are still not well understood.

The recent study by Dong *et al.*[25] represents a noteworthy step forward, as they succeeded in mapping the near-field distribution of an all-dielectric BIC metasurface by using electron energy loss spectroscopy (EELS) and cathodoluminescence (CL) based on a transmission electron microscope (TEM). However, the practical applicability of EELS and CL for near-field mapping of individual resonators and arrays of resonators is limited due to the high cost and intricate operation procedures of TEM equipment, as well as the necessity to fabricate the metasurfaces on ultra-thin membranes with thicknesses of only around 30 nm. Furthermore, investigations based on EELS and CL cannot resolve the optical phase and are constrained in fully describing the out-of-plane electric field of such structures, consequently omitting crucial information integral to the complete characterization of collective mode properties.

In order to overcome these limitations, we leverage scattering-scanning optical near-field microscopy (s-SNOM)[26,27] as an alternative state-of-the-art method for the characterisation of the near-field properties of photonic quasi-BIC metasurfaces. S-SNOM operates by focusing a laser beam on the tip of an atomic force microscope (AFM) creating an optical near-field at the apex of the tip and subsequently collecting the backscattered signal. This approach enables spatial resolution dictated solely by the tip radius[28], thus surpassing the limitations imposed by the diffraction limit. The demodulation and interferometric detection of the backscattered light allows the retrieval of both the optical amplitude and phase of the sample's near-field together with the correlated topography of the sample surface.[27] The technique has been widely applied for the investigation of the near-field response of single[29–32] and arrays[33] of plasmonic and dielectric resonators. In contrast to the limitations faced by EELS, s-SNOM does not require the metasurface to be fabricated on nanometer-thin membranes. Instead, it enables the non-destructive study of the sample fabricated in a configuration where it can be directly employed for on-chip applications.

Here, we augment the capability of s-SNOM with an algorithmic image processing technique that extracts the experimentally recorded optical near-field phase of individual resonators, an observable which is in principle independent of the laser power used. Subsequently, the method compares the experimental data with the phase obtained from numerical simulations to assess the strength of the non-local quasi-BIC mode at the individual resonator level. Through this approach, we unravel the finite array effect of quasi-BIC mode formation. Notably, our investigations reveal that a minimum size of a 10 x 10-unit cell array is sufficient for the complete formation of the quasi-BIC mode, a number significantly smaller than what our far-field measurements on the same metasurface indicate.

Furthermore, we investigate the directional coupling effects in quasi-BIC resonators chains and show that the coupling strength between individual resonators is much stronger in the direction of electric field polarization. In addition, we experimentally verify that isolated single defects of missing resonator unit cells exert only minor effects on both the far-field response and the near-field enhancement. In contrast, we find that larger defects, exemplified by 3x3 missing unit cell configuration have a substantial impact on the far-field response and lead to a directional attenuation of surrounding resonator near-fields.

Moreover, from an experimental standpoint, especially important for applications such as biosensing or catalysis, it is crucial to understand how the quasi-BIC-mode decays towards the array edges, where extended periodicity in all in-plane directions is not given. Our near-field measurement could experimentally capture the decay, unveiling a noticeable quenching of the BIC mode up to 7-unit cells distant from the array border. Our findings underscore that there is a notable discrepancy between the observed far-field response and the now easily accessible near-field response of quasi-BIC metasurfaces. We believe that this study underscores the capabilities of s-SNOM for investigating optical metasurfaces and serves as a starting point for deepening our fundamental understanding of the quasi-BIC mode. Beyond the fundamental understanding our findings bear significant practical implications for optimizing the spatial footprint and increasing the active metasurface area to enhance the performance for applications such as catalysis,[16] thermal emission[34] or lasing.[35,36]

**Results**

To demonstrate the efficacy of our approach for accurately imaging and assessing the mode formation, we analyze a widely used BIC-driven metasurface geometry composed of pairs of tilted ellipses[37] (**Figure 1a, b and c**). Based on previous reports,[24] we chose a tilting angle α of 20° which provides high far-field modulations while maintaining a comparably high Q factor of more than 80 in experiments (**Figure 1d**). The spectral line width can be tuned by changing the tilting angle α and the spectral position is modified by uniformly scaling the in-plane dimensions of the unit cell. Based on this metasurface concept, we study the effects of finite array sizes **(Figure 1e)**, the directional coupling of the resonators **(Figure 1f)**, edge effects **(Figure 1g)** and structural defects **(Figure 1h)** in the near-field, facilitated by a transmission mode s-SNOM setup using a pseudo-heterodyne detection scheme (**Figure 1 i, j**).[31,38] The investigated quasi-BIC mode is highly dispersive and shifts spectrally when the angle of incident light is changed.[39,40] To circumvent the resonance attenuation caused by this effect, the metasurface is excited with a normal-incident plane wave achieved through loosely focusing the laser beam with a parabolic mirror positioned below the sample. This configuration further minimizes tip-resonator coupling effects as the light polarization is perpendicular to the tip shaft[31]. Thus, the tip functions as a passive element which only scatters the metasurface near-fields with minimal perturbation to the quasi-BIC resonance. In order to achieve a high signal to noise ratio (SNR) and minimize unwanted background far-field signal, we analyzed the third order demodulated optical amplitude ($s_3$) and phase ($\varphi_3$) of the pseudo-heterodyne near-field signal. This demodulation order does not contain far-field background signal on the resonator (see **Figure S1** deproach curve on resonator).[38] Notably, the recorded s-SNOM signal from the resonators closely mirrors the amplitude and phase of the axial electric field $E_z$ on top of the resonator, as attested by the excellent agreement between the numerically simulated field $|E_z|$ and phase $\varphi(E_z)$ (**Figure 1 k, l**) and the recorded near-field optical amplitude ($s_3$) and phase images ($\varphi_3$) (**Figure 1 m, n**).[31] Particularly noteworthy is the characteristic quadrupole pattern in the experimental near-field phase, induced by the non-parallel opposing dipoles within resonator pairs that signifies the formation of a quasi-BIC mode.

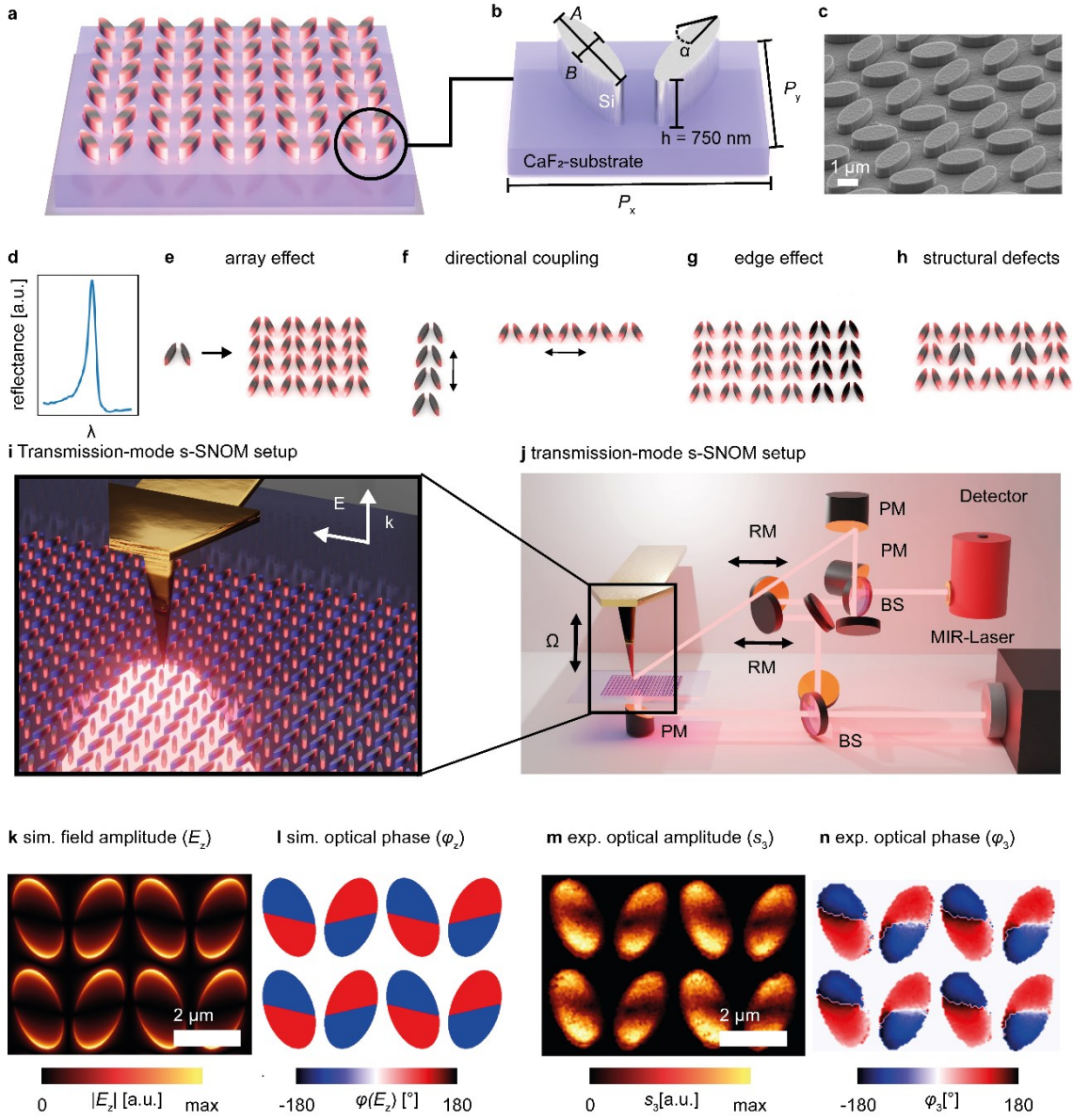

**Figure 1. (a)** Conceptual image of an all-dielectric quasi-BIC metasurface composed of pairs of tilted elliptical resonators. **(b)** A single unit cell showing a pair of tilted ellipses (α= 20°) made of silicon on a CaF$_2$ substrate. The structure is defined by the ellipses' long axis A and the short axis B and the pitch in x-direction Px and in y-direction Py. **(c)** SEM image of the fabricated metasurface. **(d)** Far-field reflection spectrum of a quasi-BIC metasurface resonance taken on **(c)**. Schematics showing an overview of the studied properties in quasi-BIC metasurfaces: **(e)** finite array size, **(f)** directional coupling, **(g)** edge effects and **(h)** structural defects. **(i,j)** Transmission mode s-SNOM used for the near-field imaging based on AFM working in tapping mode (Ω= 250 kHz) with a laser focused onto the apex of the tip by a parabolic mirror (PM). The backscattered light from the tip is collected by a second PM and superimposed on a beam splitter (BS) with a reference beam and detected by a MCT-detector. The reference beam is modulated by a vibrating reference mirror (RM) resulting in a Mach-Zehnder interferometric beam path. **(k)** Numerically simulated optical near-field amplitude $|E_z|$ and **(l)** phase $\varphi(E_z)$ of the axial electric field of a quasi-BIC metasurface unit cell. **(m)** The corresponding experimentally measured demodulated (n= 3) optical near-field amplitude ($s_3$) and **(n)** near-field phase images ($\varphi_3$) of 30x30 array of resonators.

To quantify the collective quasi-BIC mode, it is important that the chosen approach remains independent of the excitation laser power. Thus, the laser-power dependent near-field amplitude $s_3$ is inadequate for the comparative analysis of the near-field signals across different structures and samples, as a potential normalization process of the optical amplitude for each measurement is error prone and time-consuming. To overcome this constraint, we introduce a method that is based on the experimentally measured near-field phase $\varphi_3$ as a more reliable observable for comparisons between measurements on different structures, since it should, in principle, remain invariant regardless of the power utilized.

Given that s-SNOM is based on an AFM platform, it allows the correlative recording of both optical and topographical data. In our method, we first use the recorded topography $Z$ **(Figure 2a, b)** to extract the individual resonators. This involves choosing a threshold value $Z_{threshold}$ ($Z_{threshold} = 0.95 \cdot Z_{max}$ in our case) and setting all values below the threshold to 0 to separate the resonator surfaces from the surrounding topographic dataset, effectively isolating the resonators. This refined dataset can then be leveraged in combination with a *k-means* clustering algorithm,[41] a widely employed method in image processing and data analysis across diverse research areas such as biology,[42] in order to separate and identify each individual resonator (**Figure 2c, d**) and generate a grid based on the unit cell dimensions of the metasurface (**Figure 2e**).

Next, we overlayed the grid onto the recorded near-field optical phase images **(Figure 2f)**. We then quantify how closely the measured phase of each resonator matches the numerically simulated near-field phase of the axial electric field $\varphi(E_z)$ by computing a surface integral of the difference between experimental recording and the numerical simulation over the entire resonator surface (**see Supplementary Note 1**). For the recorded pixelated image, this corresponds to subtracting the measured and numerically simulated near-field phase at each pixel and obtaining a map that shows how closely $\varphi_3$ resembles the ideal phase of opposing dipoles **(Figure 2 g and h)**.

This computed score can then be averaged for each resonator pair, yielding a comprehensive map of the entire image revealing regions with a strong or weak quasi-BIC mode formation **(Figure 2i)**. We denote this figure of merit (FOM) *BIC mode purity parameter* (*B*), which, by definition, ranges from 0 to 1 and describes the extent to which the near-field phase pattern mirrors the ideal quasi-BIC phase pattern derived from numerical simulations. In the upcoming sections, we will show that this parameter can provide additional critical information only obtained through near-field measurements on quasi-BIC mode properties.

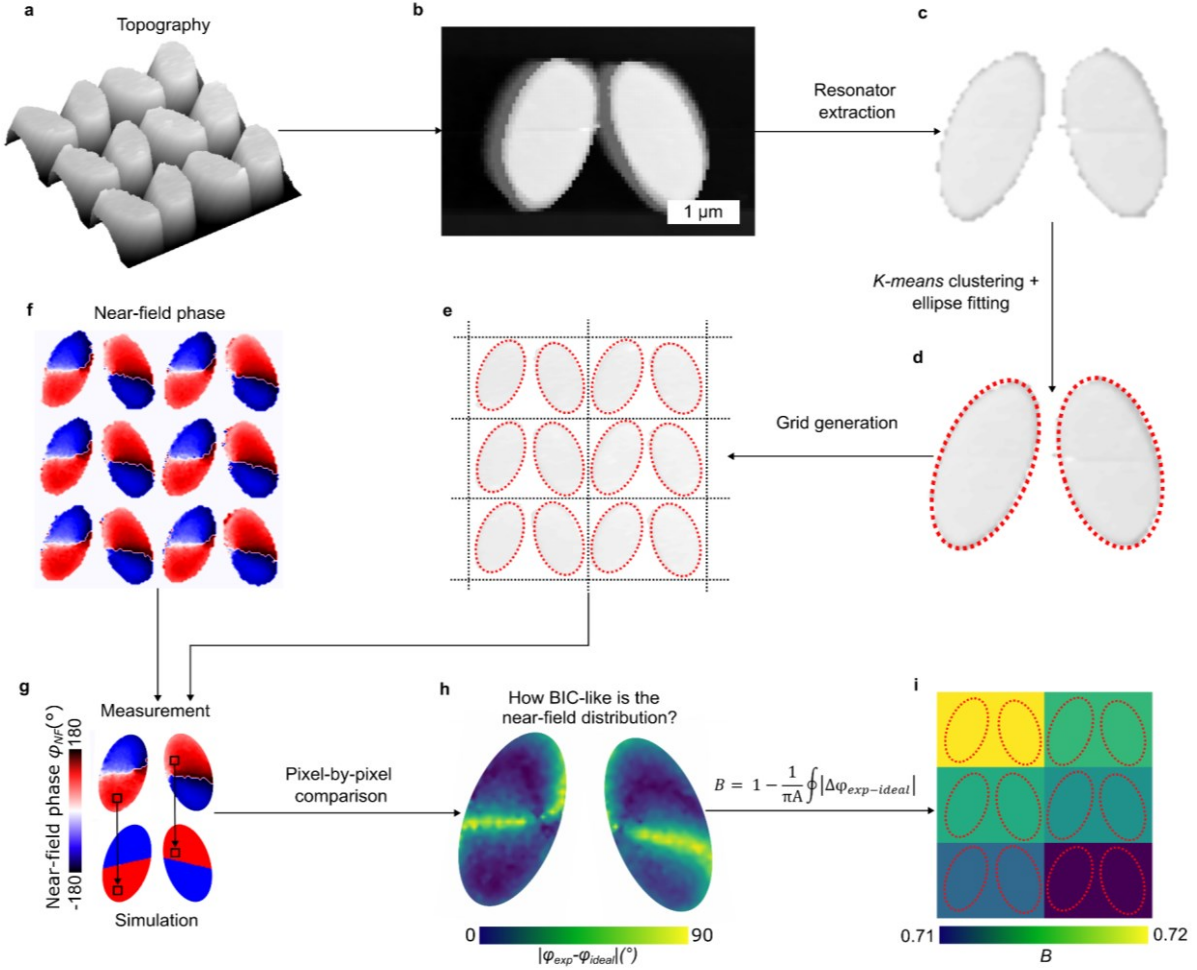

**Figure 2.** **(a)** Recorded topography $Z$ on a section of a 20 x 20 array of tilted ellipse resonators. **(b)** Topography of an individual unit cell. **(c)** Resonator pair extracted from **(b)** by setting an arbitrarily chosen threshold $Z_{Threshold}$ for the height $Z$-value (here $Z_{Threshold}$ = 0.95 $Z_{max}$). **(d)** Individual resonators are isolated by applying a clustering algorithm (*k-means)* and ellipses are fitted to each resonator. **(e)** A grid is generated based on the fitted ellipses in order to create a bounding box for each unit cell. **(f)** Correlated near-field optical phase image $\varphi_3$ recorded simultaneously to **(a)**. **(g)** Comparison between measured near-field phase $\varphi_3$ of an individual unit cell extracted via the procedure described in **(b-e)** and numerically simulated near-field phase. **(h)** Pixel-by-pixel subtraction of experimental data and simulation. **(i)** Calculation of the *BIC mode purity parameter B*, which can be described as an integral over the entire resonator surface, containing the difference of experimental and simulated near-field phase.

The first property we investigated with this newly introduced method of analysis is the impact of finite array sizes on the formation of the collective BIC mode. As fabricated metasurfaces are limited to finite array size, which perturbs the BIC mode and consequently lowers the achievable Q factor and field enhancement.[4] Although previous simulation and experimental far-field studies have been conducted on finite array size effects of quasi-mode formation,[4,43–45] a critical gap exists in the experimental exploration of the near-field behaviours of finite quasi-BIC arrays essential for example for catalytic applications. For this study, we fabricated arrays of N x N-unit cells ranging from 1 x 1 to 30 x 30 **(see SEM picture in Figure 3d, e, f and g)** and first measured each array with a standard far-field QCL-IR microscope **(see Materials and Methods)**. The maximum reflectance image and spectra of these arrays **(Figure**

**3a, b)** display a continuous increase in peak reflectance with the expansion of the array size N. Furthermore, the quality factor of the quasi-BIC resonance deduced from the reflectance spectra **(Figure 3c and Figure S2)** steadily increases with N reaching a quality factor of around 85 for a 30 x 30 array size. Noteworthy is the challenge in characterizing array field sizes smaller than 10 x 10-unit cells with standard far-field IR microscopes, as the signal cannot be reliably distinguished from background noise.

In contrast, the near-field images of the same arrays recorded at the center of each array **(shown as red squares in Figure 3d, e, f, g and in Figure S3)** and the derived BIC mode purity *B* **(Figure 3h)** show different behavior of the collective quasi-BIC mode. The calculated *B*-Score equals $B = 0.18$ for a single unit cell and exhibits a rapid increase with each successive addition of unit cells to around $B = 0.55$ for a 5 x 5 array field. This increase underscores that every unit cell addition strengthens the quasi-BIC mode. Strikingly, after N>10, the *B*-Score saturates between 0.65 and 0.75 and does not increase substantially even when the array size is increased. This trend shows that the quasi-BIC mode is already fully formed at an array size of approximately 10 x 10 resonators, a significantly smaller scale than implied by our far-field measurements. We attribute the mismatch between the observed full formation of the quasi-BIC mode on individual resonator level for N = 10 and the increase in far-field reflectance and quality factor with increasing N>10 to a higher directionality of the transmitted or reflected light towards the microscopy objective[44–47]. The discrepancy between the ideal B-score of 1.0 and the experimentally observed saturated score of 0.65 to 0.75 can be caused by several factors, such as fabrication inhomogeneities[48] and minor asymmetries in bottom illumination of the metasurface by the bottom parabolic mirror.[44]

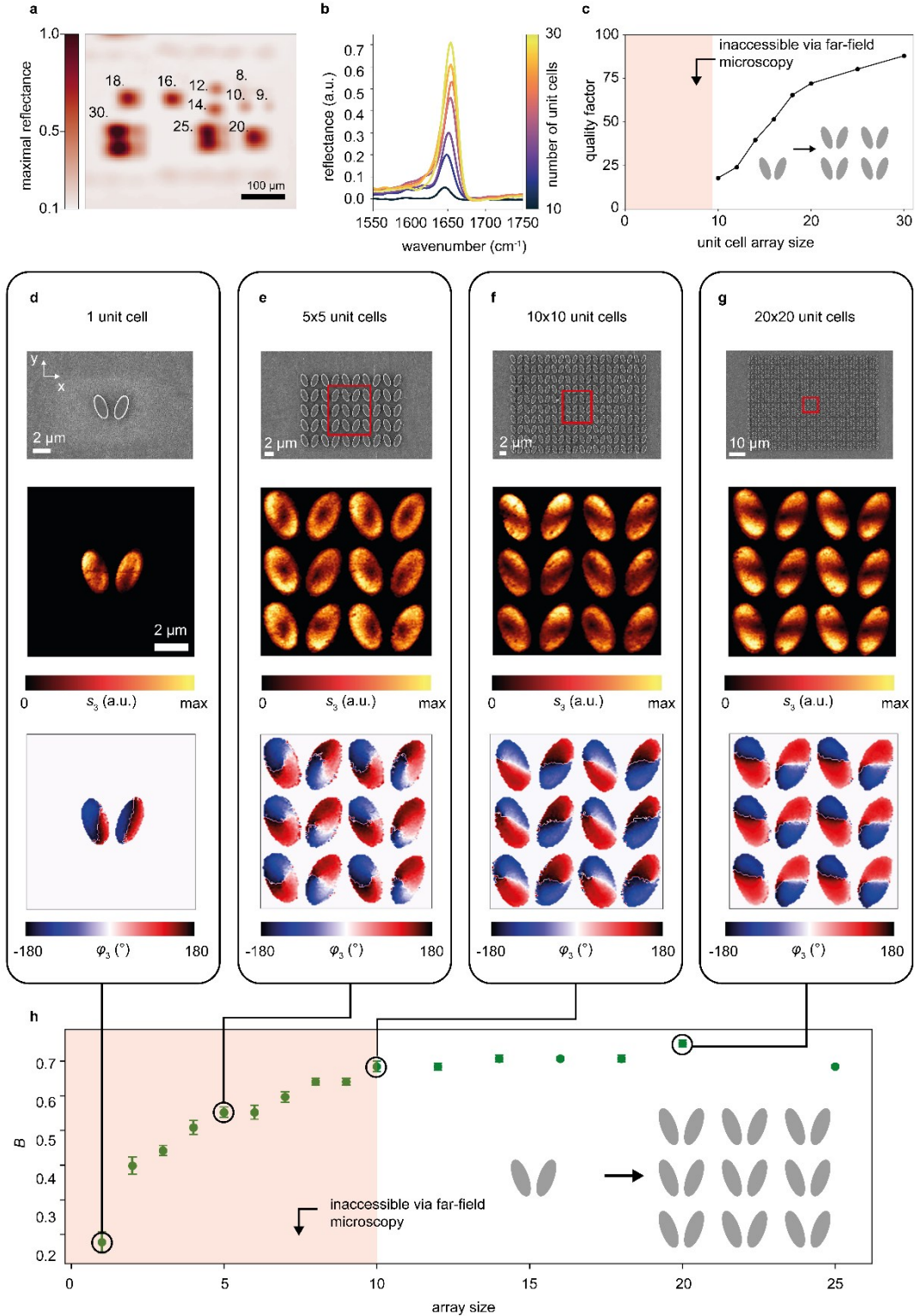

**Figure 3. (a)** Peak reflectance image of different array sizes ranging from 8 to 30-unit cells. **(b)** Reflectance spectra of the different sized arrays. **(c)** Plot of the extracted quality factors from the spectra in **(b)**. SEM, optical near-field amplitude ($s_3$) and phase images ($\varphi_3$) of different arrays with 1 **(d)**, 5x5 **(e)**, 10x10 **(f)** and 20x20 **(g)** unit cells recorded at the position of the red square in the SEM images. **(h)** Calculated *BIC* mode purity parameter $B$ from the experimental near-field phase images.

Our investigation of the finite array size effect on the near fields was predicated on the conventional quadratic arrangement of N x N-unit cells. Although this arrangement has until now been dominant in the literature[4,37,49,50], it is often based merely on intuition and convention. Here, we want to challenge this preconception and explore possibilities to further reduce the footprint of metasurfaces, while maintaining the quality of the quasi-BIC mode. For this purpose, we studied the quasi-BIC mode formation in metasurfaces with one long axis comprising 25-unit cells, and one short axis of N-unit cells, starting at N=1. Thereby, we differentiate two main design cases with one case having the long axis in x-direction aligned along the polarization direction of the electric field (25 x N) and the other case having the long axis perpendicular to the x-direction and thus perpendicular to the polarization direction (N x 25).

For this comparative analysis, we fabricated arrays based on unit cell dimension of 25 x N or N x 25 **(Figure 4a, b, c and d)**. A simple comparison of the 25 x 1 and 1 x 25 near-field distributions **(Figure 4a vs. Figure 4c,** phase images recorded at the position of the red square**)** immediately underscores a clear difference in the near-field behavior of the different design cases. Whereas the 25 x 1 array displays an emerging quasi-BIC phase pattern with $B = 0.5$, the 1 x 25 array does not exhibit the characteristic quadrupole phase pattern of two opposing dipoles and has a much lower score of $B = 0.29$. This discrepancy shows a preferred coupling direction for the 25 x N array case. The full comparison of measurements on arrays of either 25 x N, N x 25 and N x N-unit cells **(Figure 4e and Figure S4)** further underlines this disparity in mode formation. The rectangular 25 x N arrays demonstrate a fully established quasi-BIC mode with a saturation score of $B = 0.7$ for $N \geq 2$. In stark contrast, the second rectangular case of N x 25 shows a much slower increase of $B$, consistent with the collective mode behavior of the classic N x N array and attains a fully established quasi-BIC mode only for $N = 10$ **(compare Figure 3h)**. We attribute this strongly differing mode formation to a difference in the strength of the directional coupling of the resonator chains. The net dipole moments of the unit cells along the polarization direction can constructively interact, facilitating the formation of the BIC mode even with a singular chain of resonators in the 25 x 1 design case. Conversely, columns of unit cells perpendicular to the polarization cannot constructively interact, thereby necessitating an array size ($N = 10$) for quasi-BIC formation. The analogous mode formation characteristics observed between the N x N configuration and the N x 25 arrangement highlight the predominant influence of coupling along the polarization direction, while the perpendicular coupling demonstrates minimal impact on mode formation. This trend is further corroborated by far-field measurements on the same metasurfaces (**Figure 4f, S5 and S6 for the spectral data and fit**), where 25 x N arrays achieve their maximum quality factor at $N = 9$, considerably earlier than the N x 25 arrays, which require N = 25 to obtain a comparable quality factor. These findings suggest that a rectangular array design featuring a substantial number of unit cells aligned along the polarization axis can exhibit a similar quasi-BIC mode quality compared to conventional square metasurfaces, albeit with considerably reduced footprint. Moreover, this directional coupling concept provides a compelling explanation for design principles governing radial BIC geometries, where lines of BIC unit cells are arranged in a circular pattern for compactness and polarization invariance .[3]

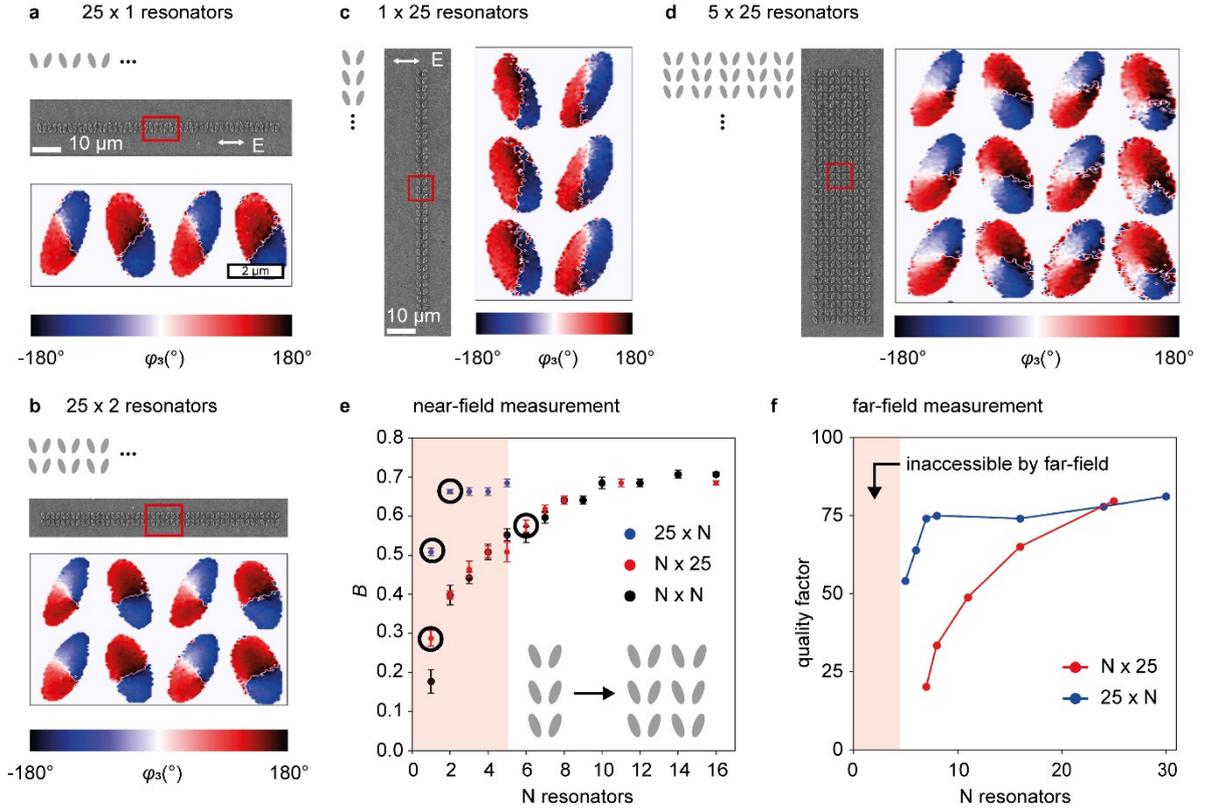

**Figure 4.** SEM and optical near-field phase images ($\varphi_3$) recorded at the position marked by the red square of **(a)** a horizontal chain of 25 x 1-unit cells and **(b)** of 25 x 2-unit cells. SEM and near-field phase images of **(c)** a vertical chain of 1 x 25-unit cells and **(d)** 5 x 25-unit cells. **(e)** Extracted BIC mode purity parameter $B$ from near-field phase images. **(f)** Extracted quality factors from far-field reflectance spectra of different numbers of columns and rows. The red squares marks the region, where far-field measurements are not able to differentiate the signal from background noise.

Besides the quenching of the quasi-BIC mode imposed by the finite array size, another attenuating effect on the quasi-BIC can be caused by lattice defects within the array. Despite their commonplace occurrence during the fabrication process, the impact of these irregularities on neighboring unit cells remains largely unexplored. Leveraging the s-SNOM technique, we can now directly quantify the influence of such irregularities on the mode formation. For simplicity, vacancies in the form of missing unit cells were chosen as perturbations of the metasurface. Various sizes of vacancies were fabricated, ranging from single- (1x1) up to multi-vacancies (5x5) in the center of a 30 x 30-unit cell metasurface array **(Figure 5 c, f)**. The resulting far-field reflectance spectra and extracted quality factors measured on these structures reveal that a solitary unit cell defect induces only marginal alterations in the far-field response **(Figure 5 a,b and S7 for the spectral data and fit)**, yielding a quality factor of $Q = 68$ compared to $Q = 69$ for the same structure without defects. However, the introduction of a 2 x 2-unit cell vacancy already substantially lowers the quality factor and peak reflectance of the metasurface to below $Q = 65$. This behavior is also seen in the near-field images of the same structure **(Figure 5 d, e)**. In these specific measurements the near-field optical amplitude ($s_3$) can be directly compared for each resonator, because the optical amplitude of all resonators are recorded in the same measurement with negligible laser power fluctuation and signal drift. Importantly, the BIC mode purity $B$ of the metasurface with a single unit cell defect does not decrease around the vacancy, and the near-field amplitude $s_3$ only shows a minimal decrease for the unit cells adjacent to the defect parallel to the polarization axis. Conversely, in the case of a 3 x 3-unit cell defect **(Figure 5 g, h)**, both $s_3$ and $B$ exhibit a substantial decrease at the periphery of the defect and along the polarization axis (to the right and left of the defect). In contrast, the unit cells neighboring the defect perpendicular to the polarization axis show a smaller decrease in $s_3$ and $B$. These measurements underscore the pivotal role played by unit cell neighbors horizontally aligned with the polarization axis in establishing the collective quasi-BIC mode, in contrast to their vertically aligned counterparts. Ultimately, these measurements show the tolerance of the quasi-BIC metasurfaces against single missing unit cell defects, both from a near- and a far-field perspective, which are the most common vacancies occurring during fabrication processes. However, their performance is significantly compromised in the presence of larger defects, particularly those exceeding the size of 2 x 2-unit cells.

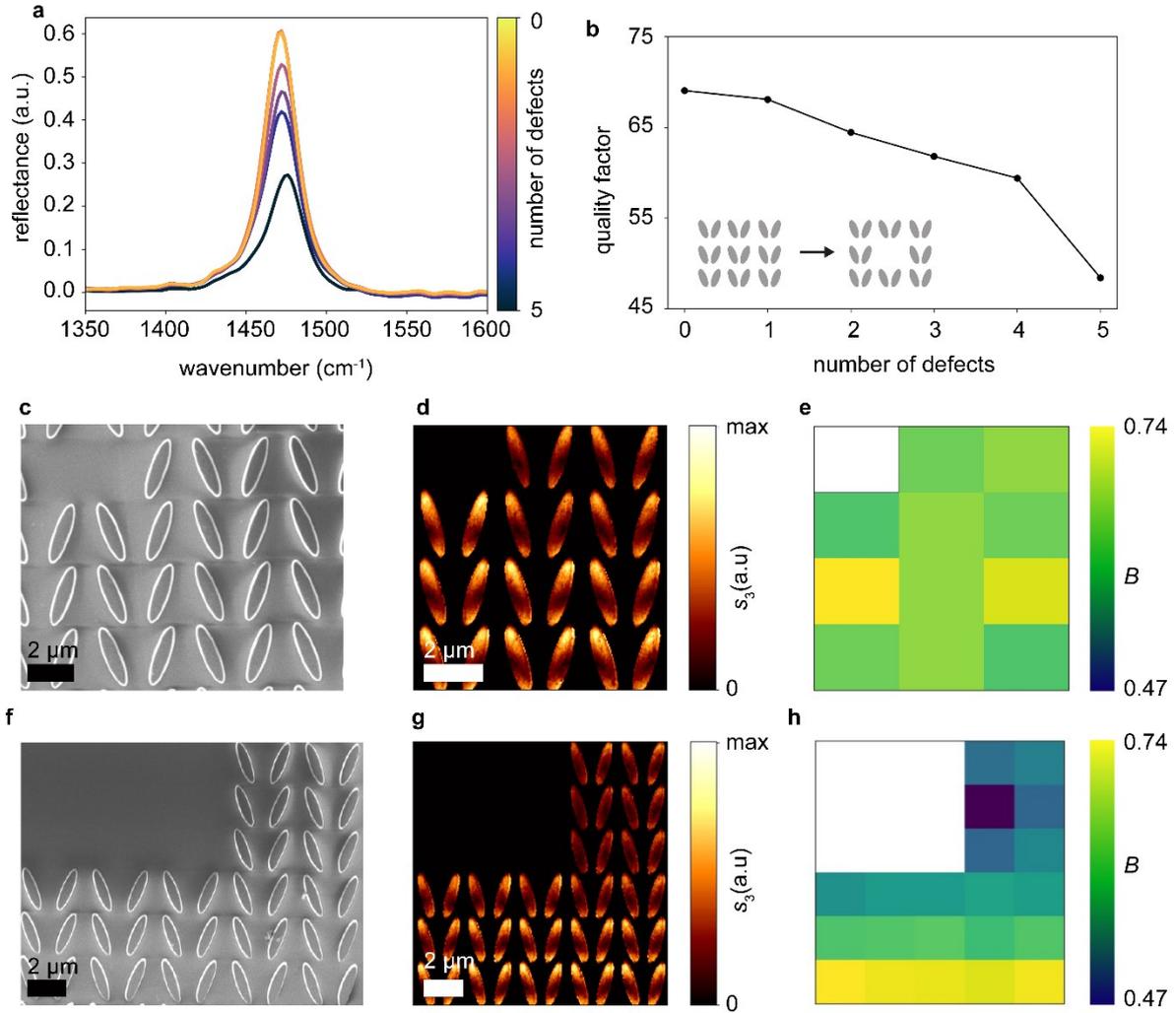

**Figure 5. (a)** Reflectance spectra of 30 x 30-unit cell array with a central defect ranging in size from 0 to 5 x 5-unit cells. **(b)** Quality factors extracted from the spectra shown in **(a)**. **(c)** SEM, **(d)** optical near-field amplitude ($s_3$) and **(e)** calculated BIC mode purity $B$ score of a 30 x 30-unit cell array with a central defect of 1 missing resonator pair. **(f)** SEM, **(g)** optical near-field amplitude ($s_3$) and **(h)** BIC mode purity $B$ score of an array with 3 x 3 missing unit cell defect.

In addition to attenuation imposed by the finite array size, edge effects are another crucial factor when determining metasurface dimensions for concrete use cases. While 10x10-unit cells are sufficient to fully create a quasi-BIC mode, high Q factors and strong field-enhancements are only present in the very center of the metasurface. However, many technologically relevant applications, such as molecular sensing, require larger areas, where the quasi-BIC mode is fully developed, to enable the averaging of spectral signals for improved signal-to-noise ratios[51,52]. Our s-SNOM method helps to decipher the negative effect of the edges and their penetration depth within the metasurface. To this end, we conducted scans along the polarization direction at the edge of a 30 x 30 array. The recorded optical near-field amplitude and the extracted amplitude at the top edge of the resonators **(Figure 6a, b)** reveal that the optical amplitude begins to decrease when approaching the seven outermost unit cells at the edge of the array.

Specifically, the optical amplitude undergoes a pronounced decline upon reaching the third resonator pair from the edge, culminating in near-complete attenuation upon reaching the final unit cell at the periphery. These findings agree well with EELS measurements of the edge effect conducted by Dong et al.[25] The near-field phase and the derived mode purity *B* (**Figure 6c, d**) show a similar trend, reaching a steady-state between 5 and 6 neighbors away from the edge, complementing the near-field amplitude measurements. Moreover, we conducted measurements at the corner of a metasurface (**Figure S7**), where we find again a pronounced susceptibility of the mode at the lateral edge, in contrast to the near-fields at the upper edge, which exhibit much higher robustness of the collective mode. This observation underlines that the horizontal coupling direction plays a dominant role, similarly to what has been observed previously in **Figure 4**. This fact supports a rectangular metasurface geometry with many unit cells aligned along the polarization axis to achieve an increase of overall active near-field area. Such a design would overall decrease the area of the metasurface susceptible to attenuating edge effects. Furthermore, these findings indicate that, when performing sensing applications or enhancing emitters for example through the Purcell-effect[53], the last six neighbors at the lateral edge of the metasurface provide a significantly reduced light-matter coupling compared to the unit cells at the center.

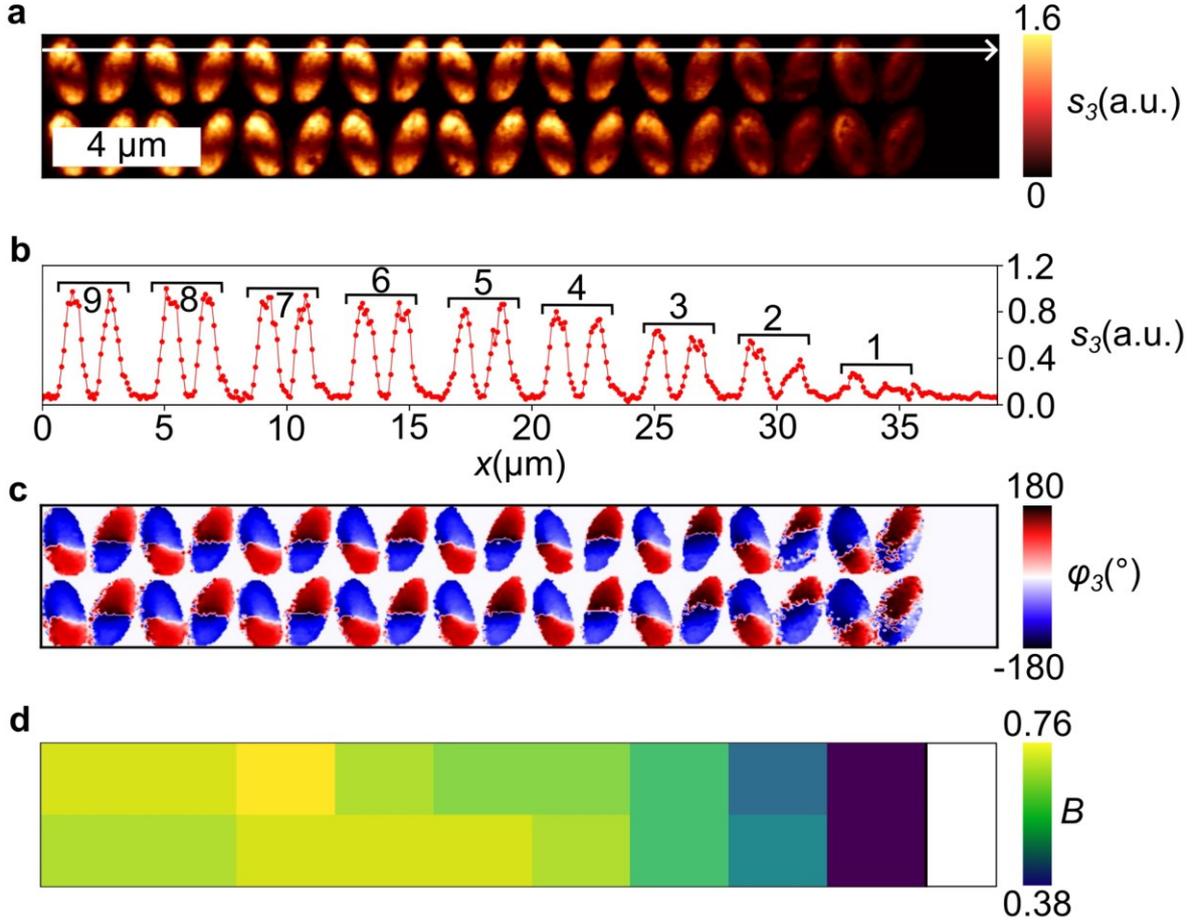

**Figure 6. (a)** Optical near-field amplitude ($s_3$) of the horizontal edge section of a 30 x 30-unit cell array. **(b)** The extracted near-field amplitude signal ($s_3$) from the white line in **(a)** with the number of unit cells marked by the number above the plotted optical amplitude of the resonators. **(c)** Optical near-field phase ($\varphi_3$) and **(d)** the extracted mode purity $B$ for each resonator pair of the same edge as in **(a)**.

## Conclusion

We have demonstrated that transmission-mode s-SNOM in combination with correlative image processing is a versatile technique to characterize photonic metasurfaces by quantifying mode properties such as the finite size effect, directional coupling, edge state effects and vacancy effects on quasi-BIC metasurfaces on the individual resonator level. We showed that the quasi-BIC mode is already established at an array size of 10 x 10-unit cells, suggesting that applications, which are solely based on the near-field of the metasurface, such as catalysis or nanoscale heating, can be performed with arrays of much smaller size than commonly assumed. We further demonstrate that commonly used square metasurface arrays are, in fact, not necessarily optimal for realizing high-Q resonances in compact nanophotonic devices. As we have shown experimentally for rectangular arrays consisting of tilted ellipses, rectangular metasurfaces with a large number of unit cells aligned along the polarization axis can deliver much smaller footprints while maintaining high mode quality and reducing the length of metasurface boundaries susceptible to edge effects. We believe our results will help to produce metasurfaces with higher active areas and smaller footprints to boost applications such as catalysis, biosensing and non-linear optics. Furthermore, we believe that our general image processing methodology for assessing mode properties from near-field microscopy data can be directly applied to a multitude of nanophotonic systems ranging from plasmonic resonators[31] and polaritons in 2D materials[54] to the study of topologically structured light such as optical vortices and optical skyrmions in thin films systems.[55]

## Methods or experimental section

**Near-field optical measurements**

Near-field microscopy measurements were conducted with the transmission pseudo heterodyne detection module of a commercial s-SNOM system (neaSCOPE from attocube systems, Haar, Germany). The laser source is an OPO-Laser (Stuttgart Instruments, Stuttgart, Germany) with a 1050 nm pump laser and tuneable MIR output obtained through DFG generation in a nonlinear crystal ranging from $\lambda = 1.4$ μm to $\lambda = 16$ μm. The output was limited by a grating monochromator to spectral width of 10 cm$^{-1}$ and a laser power of 2 mW. For the pseudo heterodyne detection,[38] the beam is split into two parts through a beam splitter. One part is focused by a bottom parabolic mirror onto the sample and the probing AFM-tip operating with tapping mode at a frequency of 250 kHz (nano-FTIR Tips, attocube systems). The backscattered light from the tip is subsequently collected by a second parabolic mirror placed above the tip and then recombined through a beam splitter with the reference beam. The second beam part is modulated through a vibrating mirror with a defined mirror amplitude to achieve a decoupling of optical phase ($\varphi_n$) and amplitude ($s_n$).[38] The recombined beam is detected by a liquid nitrogen cooled MCT-detector and demodulated based on higher orders of the tip frequency in order to eliminate the background signal. In this study, we used the third demodulation (n=3) of the recorded signal, which compromises a good trade-off between SNR and a background free near-field signal in the MIR range.

**Far-field optical measurements**

The far-field response of the fabricated metasurface was measured with a Spero MIR-QCL-microscope (Daylight solutions Inc., San Diego, USA) with a wavelength range from 950 to 1800 cm$^{-1}$ at 1 cm$^{-1}$ resolution and a 4x magnification objective with an NA of 0.15. The output beam is linearly polarized, and the recorded spectra are reference to a clean gold coated glass cover slide to eliminate the instrumental and atmospheric response.

**Numerical simulations**

Numerical electromagnetic simulations were performed with a commercially available finite-element method electromagnetic solver CST Studio Suite 2021 (Simulia, Providence, USA) with adaptive mesh refinement and periodic boundary conditions in x and y direction while open boundary conditions are used in z direction. The source of the excitation and detection are discrete port modes placed close to the unit cell. $CaF_2$ was simulated using a refractive index n of 1.4.

**Metasurface fabrication**

A 750 nm thick layer of amorphous silicon (a-Si) was deposited on a calcium fluoride ($CaF_2$) substrate by plasma-enhanced chemical vapor deposition (PlasmaPro 100 PECVD, Oxford Instruments, UK) at 180 °C. A layer of positive tone resist poly(methyl methacrylate) (PMMA) with a molecular weight of 950k was spincoated and pre-baked. To prevent charging effects during electron beam exposure, a final layer of conductive polymer (Espacer 300Z) was spin-coated on top. An electron beam lithography system (eLINE Plus, Raith GmbH, Germany) with an acceleration voltage of 20 kV and an aperture size of 15 μm was used to pattern the designed structures into the sample. The exposed resist was developed in a solution of isopropyl alcohol (IPA) and ultrapure water with a ratio of 7:3 for 60 s. The resist pattern was transferred into a hard mask consisting of 20 nm $SiO_2$ and 40 nm Cr deposited by electron beam evaporation. The liftoff was performed with a remover (Microsposit remover 1165). Reactive ion etching (PlasmaPro 100 Cobra, Oxford Instruments, UK) was used to etch the hard mask pattern into the silicon film. The remaining hard mask was removed by wet etching (Cr etch 210, NB Technologies GmbH, Germany) for the Cr layer and reactive ion etching for the $SiO_2$ layer, resulting in pure silicon nanostructures.

**Supporting information**

**Acknowledgment**

This project was funded by the Deutsche Forschungsgemeinschaft (DFG, German Research Foundation) under grant number TI 1063/1 (Emmy Noether Program) and the Center for NanoScience (CeNS). Funded by the European Union (ERC, METANEXT, 101078018 and EIC, NEHO, 101046329). Views and opinions expressed are however those of the author(s) only and do not necessarily reflect those of the European Union, the European Research Council Executive Agency, or the European Innovation Council and SMEs Executive Agency (EISMEA). Neither the European Union nor the granting authority can be held responsible for them. SAM additionally acknowledges the Lee-Lucas Chair in Physics.

**Author contributions**

T. G. and E.B. contributed equally to this work. T. G., A.A. and A.T. conceived the project. A.A. and M.B. fabricated the metasurfaces. T.G., E.B and A.M. performed the near-field and far-field measurements. E.B: developed the image processing method for data analysis. A.A. performed numerical simulations and designed the meatsurfaces. T.G. wrote the manuscript with input from all the authors. F.K., A.T. and S.A.M. managed various aspects of the project.

**Conflict of interest disclosure:**

F. Keilmann is a scientific advisor to attocube systems AG which manufactures the s-SNOM used in this study. T. Gölz obtained financial support for his PhD thesis from attocube systems AG.

**Data availability statement**

The main data supporting the findings of this study are available within the article and its Supplementary Information files. Extra data are available from the corresponding author upon reasonable request.

**Supporting Information:**

# Supplementary information

# Revealing mode formation in quasi-bound states in the continuum metasurfaces via near-field optical microscopy

*T. Gölz[1,†], E. Baù[1,†], A. Aigner[1], A. Mancini[1,2], M. Barkey[1], F. Keilmann[1], S. A. Maier[3,4], A. Tittl[1*]*

1. Chair in Hybrid Nanosystems, Nanoinstitute Munich and Center for Nanoscience (CeNS), Faculty of Physics, Ludwig-Maximilians-Universität München, 80539 Munich, Germany.

2. Centre for Nano Science and Technology, Fondazione Istituto Italiano di Tecnologia, Via Rubattino 81, Milan, 20134 Italy

3. School of Physics and Astronomy, Monash University, Clayton, Victoria 3800, AUS

4. Department of Physics, Imperial College London, London SW7 2AZ, U.K.

†These authors contributed equally

Email: andreas.tittl@physik.uni-muenchen.de

a.)

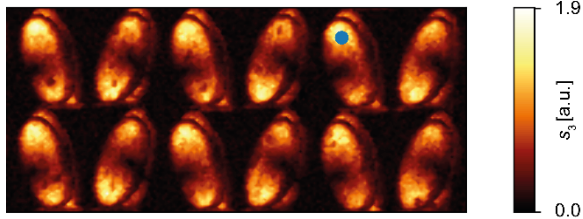

b.)

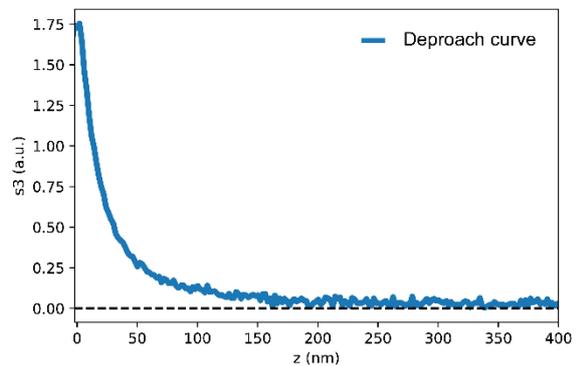

**Figure S1: (A)** Third demodulated optical amplitude scan $s_3$ taken at the center of a quasi-BIC metasurface. **(B)** Deproach curve take on the tip of an elliptical resonator (blue point in **A**) showing a drop off of the optical amplitude with increasing distance of the tip to the resonator.

**Supplementary Note 1 - Details on the BICness clustering algorithm**

The FOM introduced in this work to describe the behavior exhibited by the collective BIC mode, which we named BIC mode purity parameter *B*, is a measure of how closely the measured demodulated near-field phase resembles the ideal case of a quadrupolar phase pattern, where for each unit cell, two opposing dipoles form in each individual resonator. The fully written formula is as follows:

$$B = \frac{1}{NB_{sim}} \sum_{k=1}^{N} [1 - F_k] \in [0,1]$$

where *N* denotes the number of resonators, $B_{sim}$ is a normalization factor extracted through simulations and $F_k$ consist of surface integrals, which can be written as:

$$F_k = F_k^+ + F_k^- = \frac{1}{\pi A_k^+} \int_0^{A_k^+} dA' \left|\varphi_{k,exp}(A') - \frac{\pi}{2}\right| + \frac{1}{\pi A_k^-} \int_0^{A_k^-} dA' \left|\varphi_{k,exp}(A') + \frac{\pi}{2}\right|$$

These two integrals calculate the difference of the measured near-field phase and an ideal phase pattern, where the phase on the top surface of the resonator is shifted by 180° compared to the bottom surface. $A_k^+$ and $A_k^-$ are the top and bottom surfaces of each resonator respectively. To take into account the results yielded by numerical simulations, which do not exactly follow this pattern, the score is divided by a normalization factor $B_{sim}$, which was obtained by comparing simulations to the case of perfect opposing dipoles. For all cases, our structures yield around $B_{sim} \approx 0.9$. If this score was applied to other structures, this normalization value would, of course, have to be adjusted.

In the case when taking a scan with s-SNOM, the image would be pixelated, which changes the above mentioned integrals to summations and allows to rewrite the formula in the following way (for unit cell of length $L \times H$, assuming the first resonator has a top surface with a near-field phase of +90°):

$$B = \frac{1}{NB_{sim}} \sum_{k=1}^{N} \left[1 - \frac{2}{\pi LH} \sum_{l=0}^{L} \sum_{h=0}^{H/2} \left|\varphi_{k,exp}(l,h) - (-1)^{k-1}\frac{\pi}{2}\right| - \frac{2}{\pi LH} \sum_{l=0}^{L} \sum_{h=H/2}^{H} \left|\varphi_{k,exp}(l,h) + (-1)^{k-1}\frac{\pi}{2}\right|\right]$$

This expression can be directly applied for image processing purposes into external software such as Python or Matlab. In practice, this formula sums up the difference between experimental near-field phase and ideal case at each pixel, then sums all pixels up and calculates the average value. The sums need to remain separated, since the resonator is, as before, split into two halves, with one having an expected phase of +90° and the other of -90°. The additional terms $(-1)^{k-1}$ and $(-1)^k$ capture the behavior of the alternating sign of the near-field phase between resonator pairs. In other words, if one resonator has a top surface of +90°, the other has, in an ideal case, a top surface of -90°.

To account for phase offsets, which occur naturally during the measurements, a phase delay can be chosen in such a way as to maximize the mode purity parameter *B*. This not only yields a higher sensitivity for the FOM, but also allows for adding the evaluated offset to experimental images of the near-field phase, making them appear visually closer to a quadrupolar pattern.

The execution of the unit cell clustering and subsequent grid generation was done in a straightforward way using Python. The modules used to achieve this are "sklearn.cluster" and "sklearn.preprocessing". The ellipse fitting was conducted using the module "cv2".

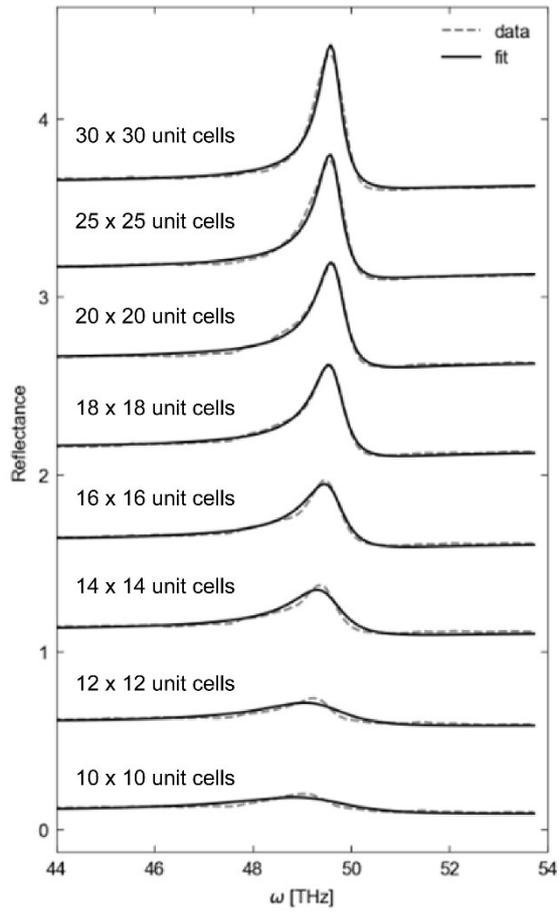

**Figure S2:** Infinite array size far-field microscopy spectra (ranging from 30 to 8) with the associated fit to extract the quality factors shown in **Figure 3c**.

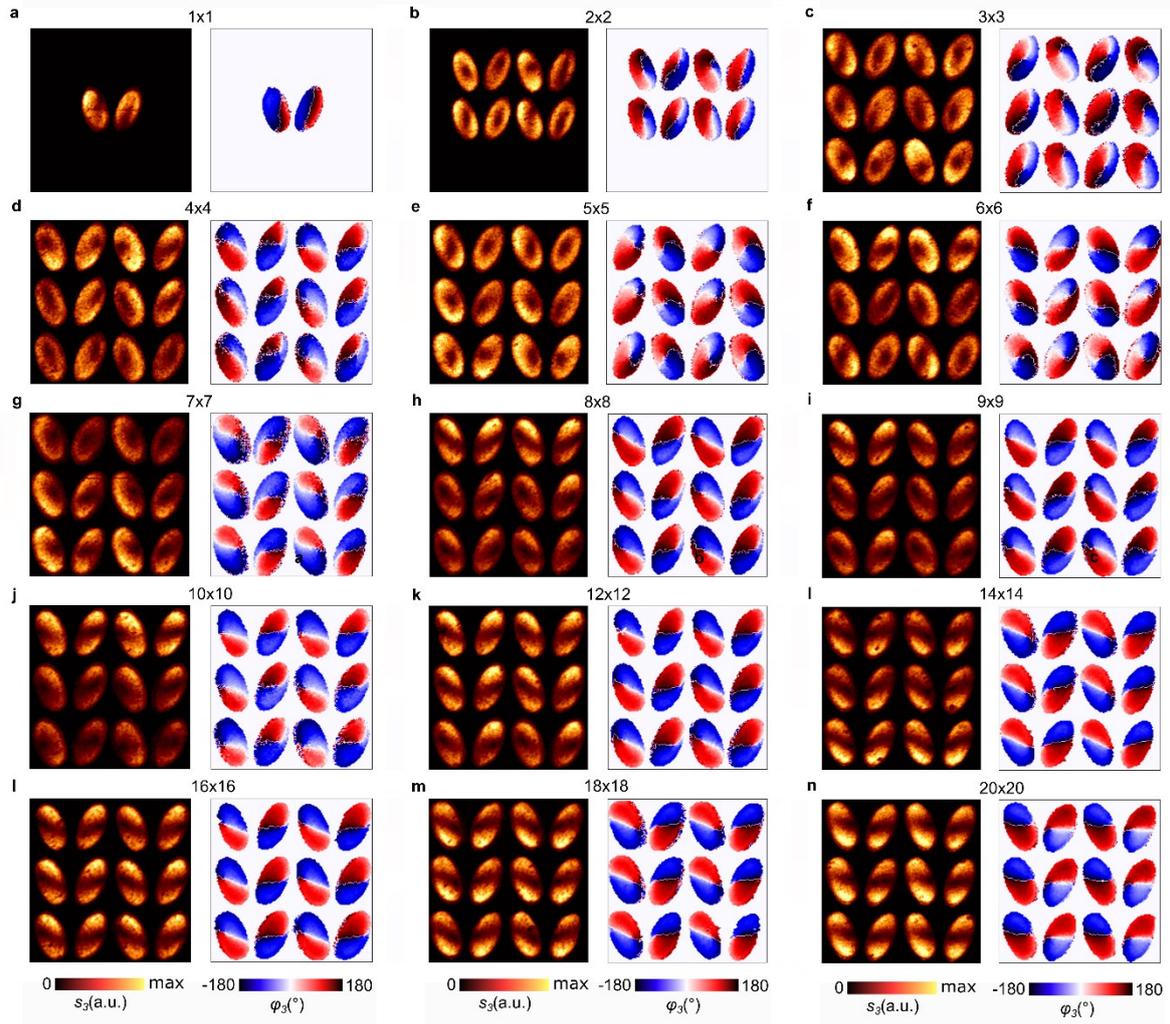

**Figure S3:** Near-field optical amplitude ($s_3$) and phase ($\varphi_3$) images recorded on arrays of different unit cell sizes used to determine the BIC mode purity parameter shown in **Figure 3h**.

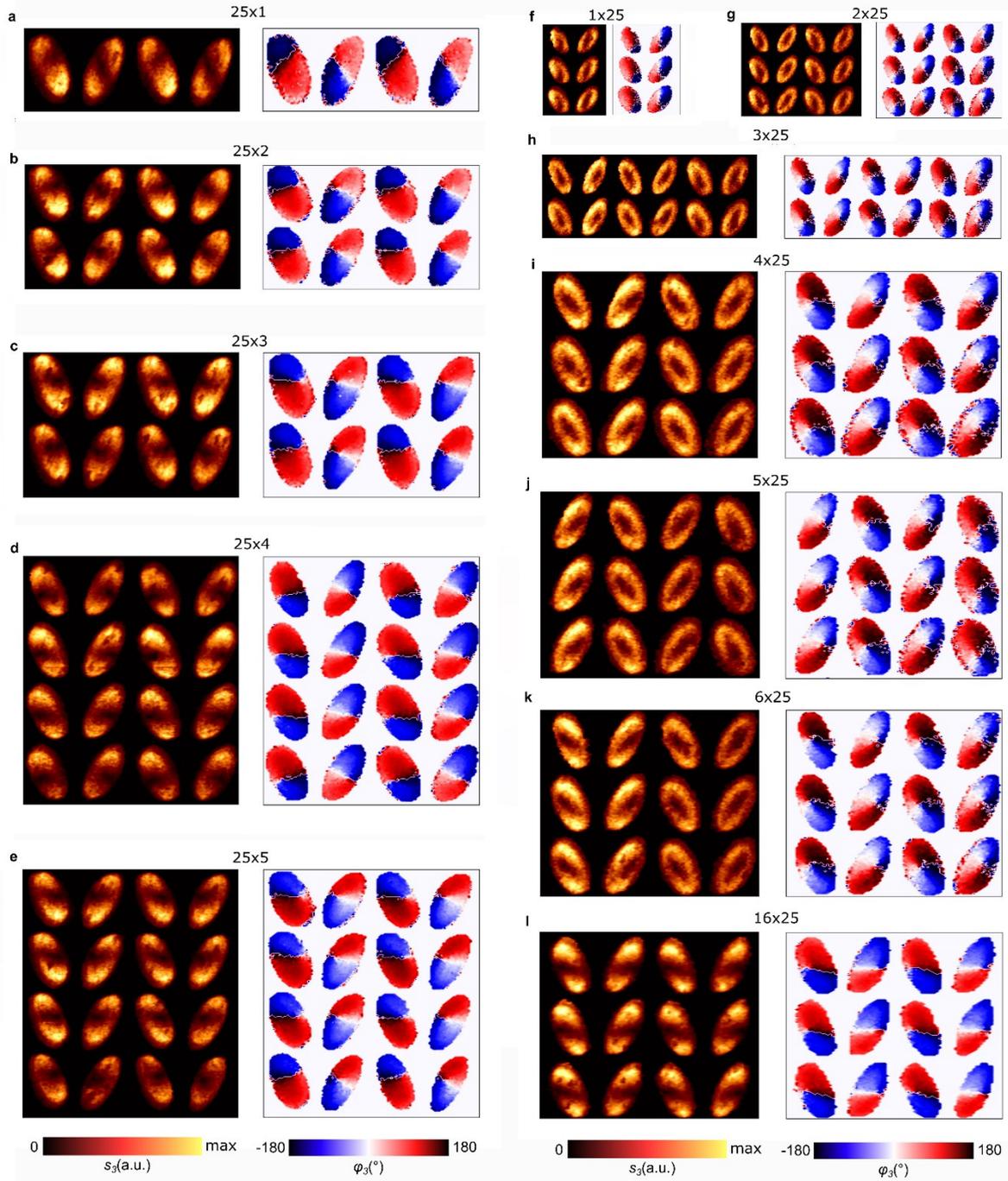

**Figure S4:** Near-field optical amplitude ($s_3$) and phase ($\varphi_3$) images recorded on arrays of different unit cell sizes used to determine the BIC mode purity parameter shown in **Figure 4e**.

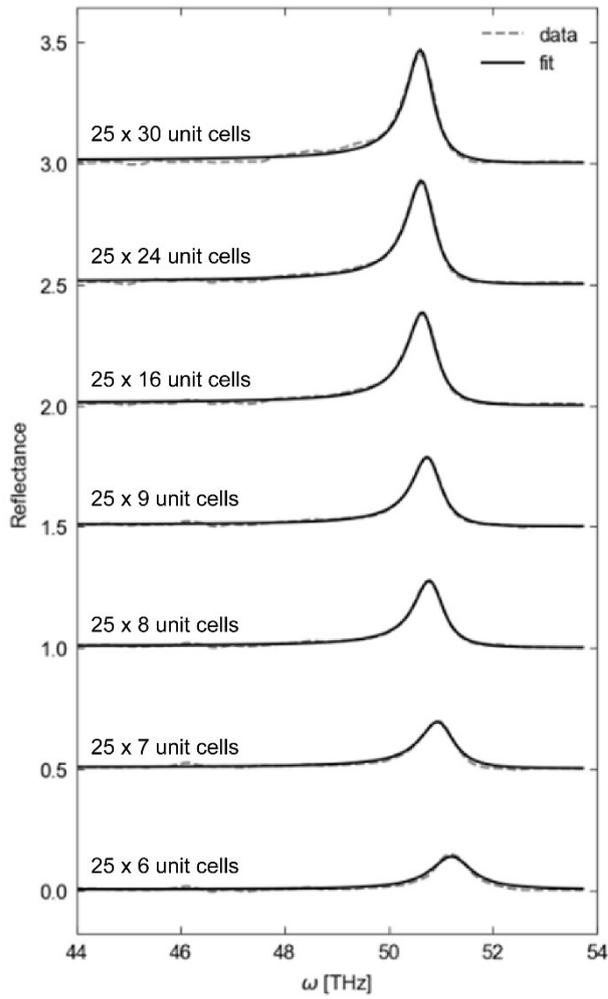

**Figure S5:** 25 x N array metasurface far-field microscopy spectra (ranging from N= 6 to 30) with the associated fit to extract the quality factors shown in **Figure 4f**. The 25 long chain of unit cells are aligned along the electrical polarization.

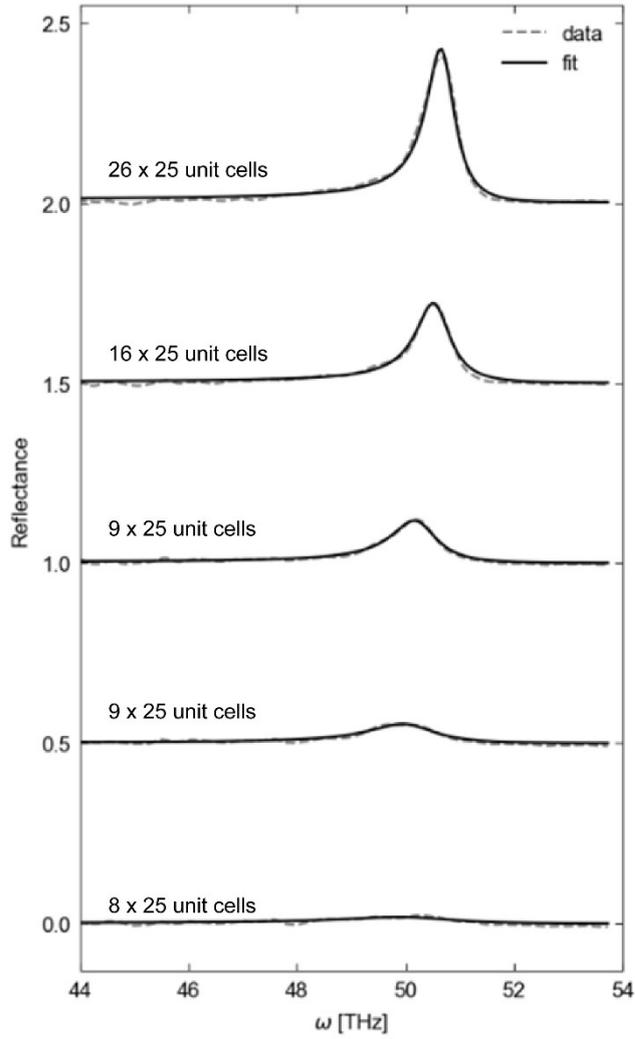

**Figure S6:** N x 25 array metasurface far-field microscopy spectra (ranging from N= 6 to 30) with the associated fit to extract the quality factors shown in **Figure 4f**. The 25 long chain of unit cells are aligned along the electrical polarization.

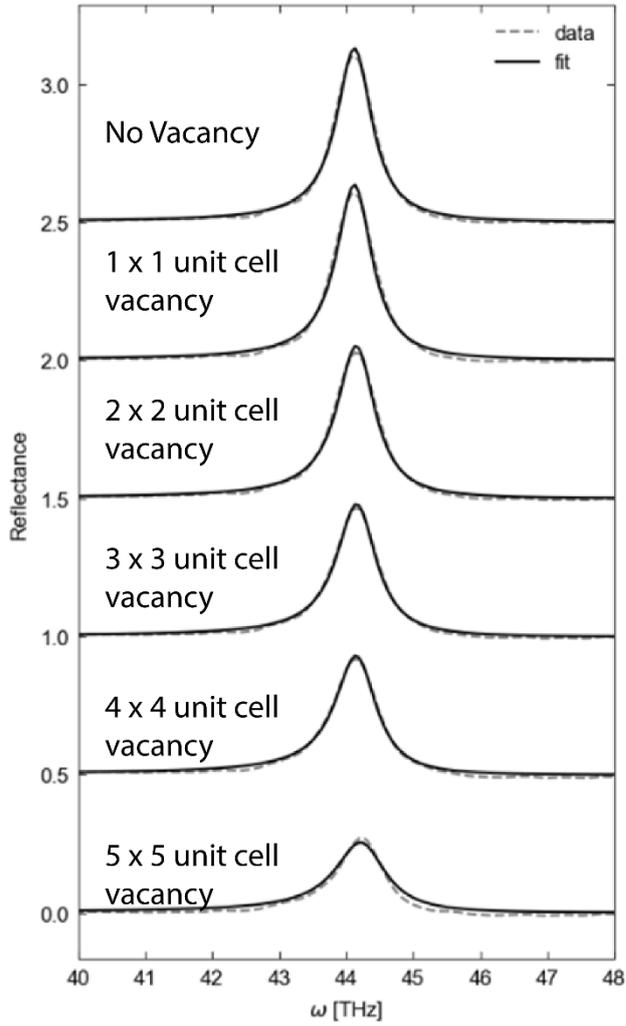

**Figure S7:** Unit cell vacancies defect far-field microscopy spectra (ranging from 0 defect at the top to 5 x 5-unit cell vacancies) with the associated fit to extract the quality factors shown in **Figure 5b**.

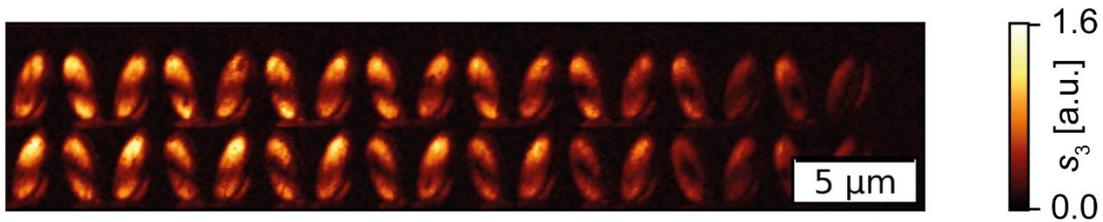

**Figure S8:** Third demodulated optical amplitude scan $s_3$ taken on the corner of a metasurface. The scan shows a decrease of the near-field in horizontal direction along the polarization axis of the electric field but no decrease in near-field intensity along the vertical direction.